\documentclass[11pt]{article}
\usepackage[margin=1in,footskip=0.25in]{geometry}
\usepackage{graphicx}
\usepackage[noadjust]{cite}

\def\ref@jnl#1{{\jnl@style#1}}
\def\ref@jnl#1{{#1}}
\def\aj{\ref@jnl{Astronomical Journal}}       
\def\actaa{\ref@jnl{Acta Astronomica}}     
\def\araa{\ref@jnl{Annual Review of Astronomy and Astrophysics}}      
\def\apj{\ref@jnl{The Astrophysical Journal}}                
\def\apjl{\ref@jnl{Astrophysical Journal Letters}}                
\def\apjs{\ref@jnl{Astrophysical Journal Supplement}}   
\def\aap{\ref@jnl{Astronomy and Astrophysics}}               
\def\aapr{\ref@jnl{Astronomy and Astrophysics Reviews}}    
\def\mnras{\ref@jnl{Monthly Notices of the Royal Astronomical Society}}  
\def\na{\ref@jnl{New Astronomy}}             
\def\prd{\ref@jnl{Physical Review D}}  
\def\prl{\ref@jnl{Physical Review Letters}} 
\def\pasp{\ref@jnl{Publications of the Astronomical Society of the Pacific}} 
\def\ssr{\ref@jnl{Space Science Reviews}} 
\def\nat{\ref@jnl{Nature}} 
\def\physrep{\ref@jnl{Physics Reports}}

\begin{document}

\title{Diverse Features of the Multi-wavelength Afterglows of Gamma-ray Bursts: Natural or Special?}

\author{J. J. Geng, Y. F. Huang\thanks{hyf@nju.edu.cn} \\
School of Astronomy and Space Science, Nanjing University,  \\
Nanjing 210046, China; \\
Key Laboratory of Modern Astronomy and Astrophysics (Nanjing University), \\
Ministry of Education, China
}
\maketitle

\begin{abstract}
The detection of optical re-brightenings and X-ray plateaus in the afterglows of gamma-ray bursts (GRBs)
challenges the generic external shock model.
Recently, we have developed a numerical method to calculate the dynamic of the
system consisting of a forward shock and a reverse shock.
Here, we briefly review the applications of this method in the afterglow theory.
By relating these diverse features to the central engines of GRBs,
we find that the steep optical re-brightenings would be caused by the
fall-back accretion of black holes, while the shallow optical re-brightenings
are the consequence of the injection of the electron-positron-pair wind
from the central magnetar.
These studies provide useful ways to probe the characteristics of GRB central engines.
\end{abstract}

\section{Introduction}
\label{sec:intro}
It is believed that gamma-ray bursts (GRBs) are generated from either
the collapse of massive stars \cite{Woosley93,Popham99,MacFadyen01,Narayan01}
or the merger of neutron stars (NSs) \cite{Eichler89,Paczynski91,Grindlay06},
during which collimated relativistic outflows can be launched. As the outflow propagates
into the circum-burst medium, a relativistic blast wave will develop, whose
dynamic evolution can be well described by the Blandford-McKee solution \cite{Blandford76}.
The blast wave would sweep-up and accelerate the circum-burst electrons
and generate afterglows at frequencies ranging from X-rays to radio waves \cite{Piran93,Meszaros97,Sari99}.
This is the basic picture for GRB afterglows.
In the pre-{\it Swift} era, many afterglow lightcurves showed a smooth
power-law decay, which can be explained by the synchrotron radiation from
electrons accelerated by the forward shock (FS).
For a complete reference of the analytical synchrotron external shock afterglow models,
one can see \cite{Gao13}.
However, many unexpected features in the afterglows were later observed
thanks to the {\it Swift} satellite \cite{Gehrels04,Burrows05,Gehrels09} and other optical telescopes
(e.g., GROND telescope, see \cite{Greiner08}).

Early flares and shallow decay phase (or the so called X-ray plateau) are
common in the X-ray afterglow data \cite{Zhang06}. This indicates that the central engines
of GRBs are still active after the burst, giving us a useful clue to investigate
the central engines. On the other hand, some optical afterglows show
re-brightenings at late stages in the observer frame ($t_{\rm obs} \sim 10^4 - 10^5$ s).
In some cases, bumps in X-rays are accompanied by optical re-brightenings (e.g.,
GRB 120326A, \cite{Melandri14,Hou14}),
while in other cases, no clear counterpart features  are observed in optical
(e.g., GRB 100814A, \cite{Pasquale15}).
Both the X-ray plateaus and the optical re-brightenings can not be explained
in the framework of a simple FS scenario.
Thus researchers have proposed several refined models to interpret these
unexpected features in recent years (see \cite{Gao15,Kumar15} for a review).
Stimulated by these refined models, it is urgent to answer
whether X-ray plateaus and optical re-brightenings have a natural
origin or they are special outcomes varying in different GRBs.

Normally, the energy released during the X-ray plateau is several percent
of the prompt emission \cite{Li15}, which motivates researchers to favour the scenarios
involving energy injection processes. According to the composition of
the injected late outflow, there are generally three types of energy injection
processes, i.e., the pure Poynting-flux injections \cite{Dai98,Zhang01,Fan06b,Kong10a},
the collision of kinetic-energy dominated shells \cite{Zhang02},
and the injection of the electron-positron-pair winds ($e^+e^-$ winds \cite{Dai04,Yu07b}).
If one further considers the optical re-brightenings, some other scenarios
are called for, including the circum-burst density jumps \cite{Lazzati02,Dai03,Nakar03},
two-component jets \cite{Berger03,Huang04},
and varying microphysical parameters \cite{Kong10b}.
All these models have succeeded to some extent in explaining one or several afterglows
according to previous studies.
On the other hand, most researchers believe that the central engines of GRBs are
either black holes (BHs) or magnetars.
Therefore, it may be reasonable to deduce that some specific groups
of afterglows should have common features, and these features are associated
with the physics of the central engines.

According to previous researches, late activities of BHs may be sustained by
the accretion of fall-back material that fails to escape from
the progenitor star \cite{Perna06,Kumar08,Wu13}. The energy injection is expected to be
delayed by the fall-back time $t_{\rm fb}$.
If the FS is affected by such a delayed energy injection,
the shock dynamics should rapidly evolve from a non-injection phase to
an injection-dominated phase \cite{Geng13}.
As a result, afterglows with steep optical re-brightenings (with the time scale
of the re-brightening $\delta t_{\rm obs} < t_{\rm obs}$) are generated.
We thus proposed that steep optical re-brightenings are
caused by the fall-back processes of central accreting BHs.
By contrast, the energy flow from a magnetar may be in the form of a
continuous $e^+e^-$ wind \cite{Geng16}.
The $e^+e^-$ wind model was initially proposed to explain the X-ray plateau \cite{Yu07a}.
The end time of the plateau phase is roughly the typical spin-down
timescale ($T_{\rm sd}$) of the newly born magnetar.
For the broad and shallow optical re-brightening, its peak time also coincides
with $T_{\rm sd}$, which motivates us to believe that the $e^+e^-$ wind
model should work for afterglows with both a shallow optical re-brightening
and an accompanied X-ray feature.

We have developed a semi-analytic method to solve the dynamic of a system
including a FS and a reverse shock (RS). It can be applied in different situations
such as when a density jump medium or the $e^+e^-$ wind is involved.
In this review, we briefly describe our related studies in recent years,
and show how the investigations help to shed light on the nature of GRBs.
In Section 2, we revisit the circum-burst density jump scenario,
and compare our results with previous hydrodynamic simulations
\cite{Eerten09,Gat13}.
The delayed energy injection model is discussed in Section 3.
In Section 4, we show the $e^+e^-$ wind model would
naturally produce the optical re-brightenings and some characteristics
of the central magnetar may be derived by comparing the theoretical results
with observations. Finally, our conclusions are summarized in Section 5.

\section{Density Jump Scenario}
After the prompt emission of GRB, a FS will form and propagate into
the circum-burst medium. The dynamic of the FS can be described
by a set of differential equations proposed by \cite{Huang99,Huang00b,Peer12}.
Assuming the number density profile of the circum-burst medium
is a step function of radius $R$: $n = n_0$ ($R < R_0$) and
$n = n_1$ ($R \geq R_0$), where $R_0$ is the transition radius
and $n_1 > n_0$.
Before the FS reaching $R_0$, the evolution of the Lorentz factor of the FS
($\Gamma_2$) is given by \cite{Geng14}
\begin{equation}
\displaystyle{\frac{d \Gamma_2}{d m_2}=-\frac{4(\Gamma_2^2-1)}
{8(1-\varepsilon_2)\Gamma_2m_2+3\varepsilon_2m_2+3M_{\rm{ej}}}},
\end{equation}
where $M_{\rm ej}$ is the initial mass of the outflow,
$m_2$ is the mass of the ambient medium swept-up by the FS,
$\varepsilon_2$ is the radiation efficiency of the shocked material. 
The subscript ``2'' is used to mark quantities in the shocked region.

When the FS encounters the density jump at $R_0$, a RS
will form and propagate back into the hot shell \cite{Sari95,Kobayashi99}.
These two shocks (FS and RS) and the contact discontinuity will
separate the system into four regions: (1) unshocked high-density medium,
(2) forward-shocked high-density medium, (3) reverse-shocked hot shell,
and (4) unshocked hot shell. In this paper, quantities in Region ``$i$'' are
denoted by subscripts ``$i$''.
We extend the derivation of \cite{Huang99} to
include the role of the reverse shock.
Firstly, it is assumed that the Lorentz factor of Region 2 ($\Gamma_2$) and Region 3
($\Gamma_3$) are equal, i.e., $\Gamma_2 = \Gamma_3 = \Gamma$.
Secondly, we can calculate the energy of each region $E_i$, and
the total energy $E_{\rm tot} = \sum_{i = 2}^{4} E_i$.
For the mass increment of Region 3, $d m_3$,
the radiative energy of the FS-RS system is $d E_{\rm rad}$.
By equating $d E_{\rm tot}$ with $d E_{\rm rad}$ and using
some additional equations, we can obtain
\begin{equation}
\displaystyle{\frac{d \Gamma_2}{d m_2}=-\frac{\frac{4}{3}(\Gamma_2^2-1)+
f_1\displaystyle\frac{d m_3}{d m_2}+
(1-\varepsilon_3)f_2\Gamma_2\Gamma_{42}(1-\displaystyle\frac{\beta_{42}}{\beta_4})m_3\displaystyle\frac{d \psi_4}{d m_2}}
{\frac{8}{3}(1-\varepsilon_2)\Gamma_2m_2+\varepsilon_2m_2+(1-\varepsilon_3)f_3m_3+\varepsilon_3m_3}},
\end{equation}
where $\Gamma_{42}$ ($\beta_{42}$) is the relative Lorentz factor (velocity)
of Region 4 as measured in the rest frame of Region 2,
$\beta_4$ is the velocity of Region 4,
and $f_1$, $f_2$, $f_3$, $\psi_4$ are functions of other
variables (see Appendix A of \cite{Geng14}).
$\Gamma_2 (R>R_0)$ could be finally derived with
some other equations involving $m_2$, $m_3$, $\Gamma_4$ and $R$.

Here, we briefly describe how the flux densities are
calculated after solving Equations (1) and (2).
Customarily, the energy distribution function, $d N_e^{\prime} / d \gamma_e^{\prime}$,
of the shocked electrons is taken as the two-segment power-law form \cite{Sari98},
with the indices of $-p$ and $-p-1$.
Hereinafter, we use prime~($\prime$) on variables to denote
quantities in the shock comoving frame and characters
without a prime to denote quantities in the observer frame.
Synchrotron and inverse Compton (IC) radiation are
then considered to calculate the emission from electrons.
Basic formulation can be found in \cite{Rybicki79,Huang00b,Fan06a,Wang10}.
Finally, the observed flux densities are obtained by integrating
emission from electrons on the equal arrival time surface \cite{Waxman97,Granot99}.

According to some previous studies \cite{Lazzati02,Dai03,Nakar03},
the optical re-brightening is attributed to the radiation from Region 3.
However, results from our refined calculations are different.
In our work, we find the magnitude of the emission from Region 3
is mainly determined by two factors.
One is the thermal Lorentz factor of baryons in Region 3, i.e, $\Gamma_{43}$
($\Gamma_{43} = \Gamma_{42}$).
The other factor is the number density of electrons in Region 3, $n_3^{\prime}$.
We set the initial values of the outflow parameters as: the
isotropic kinetic energy $E_{K,{\rm iso}} = 10^{53}$ erg,
the initial Lorentz factor $\Gamma_{2,0} = 300$,
$n_0 = 1$~cm$^{-3}$, $R_0 = 3.4 \times 10^{17}$ cm, and the redshift $z = 1$.
In Figure 1, we calculate two cases with different density jump ratios ($n_1/n_0$),
of 10 and 100 times, respectively.
Results from our semi-analytic method show that $\Gamma_{43}$
given by the analytical solution is overestimated (see Figure 1).
Moreover, we use the co-moving volume of Region 3 to calculate the
volume-averaged $n_3^{\prime}$, which is significantly lower than
that predicted by the shock jump conditions (see Figure 2 of \cite{Geng14}).
As a result, the radiation from Region 3 is 
actually lower than that given by previous analytical studies.
Figure 2 shows the corresponding lightcurves in two cases.
No notable bumps emerge after the density jump in these cases.
This is consistent with the results from several hydrodynamic simulations \cite{Eerten09,Gat13}.
In our calculations, typical values are adopted for parameters
of the plasma in all regions \cite{Huang00a},
i.e., the equipartition parameter for electron energy $\epsilon_e = 0.1$,
the equipartition parameter for magnetic field energy $\epsilon_B = 0.01$,
the electron distribution index $p = 2.3$, and the half-opening angle $\theta_j = 0.1$.

\begin{figure}
   \begin{center}
   \includegraphics[scale=0.5]{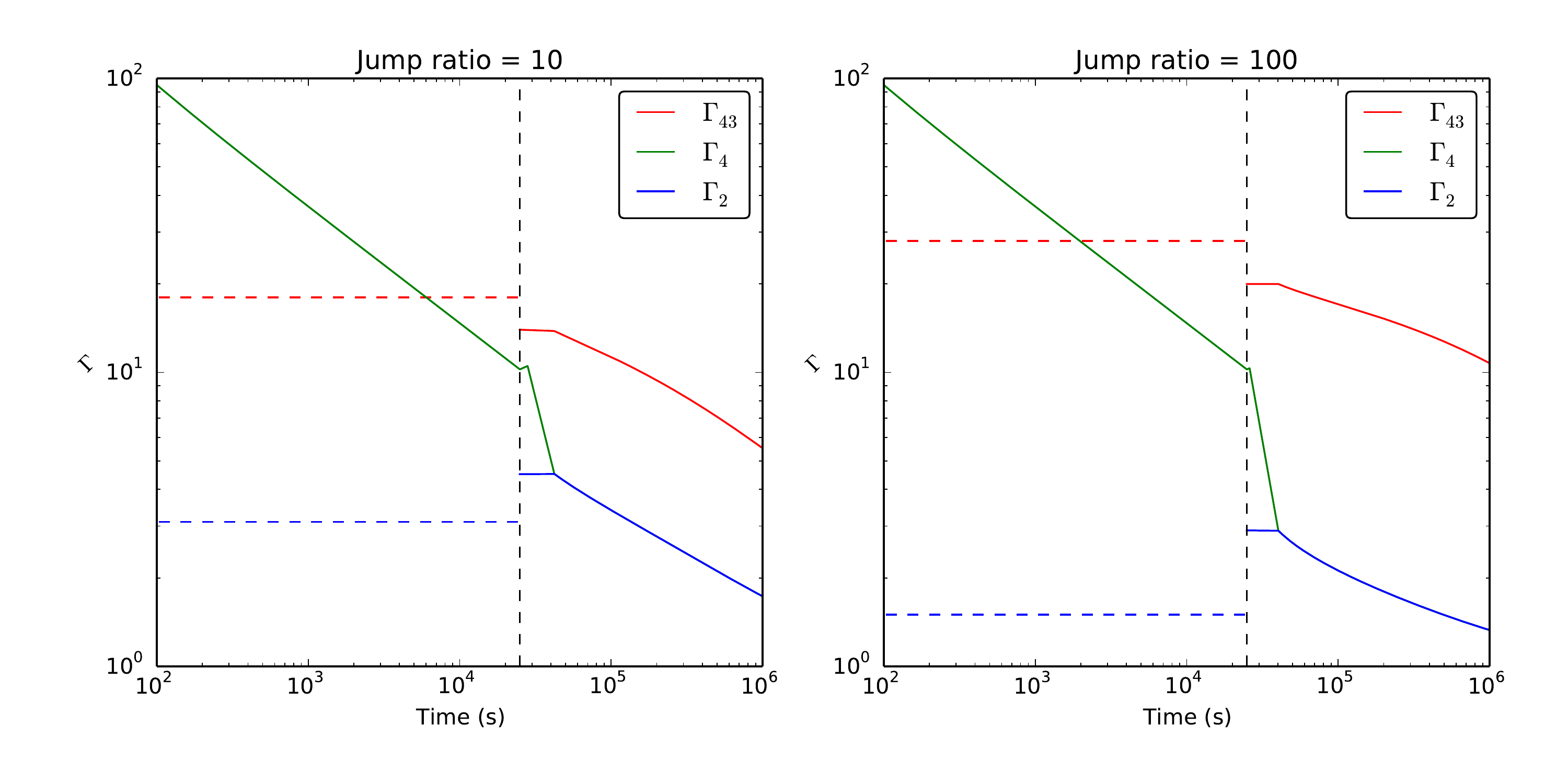}
   \caption{Evolution of the Lorentz factors of different Regions when there is a density jump \cite{Geng14}.
   Two cases are calculated, one corresponds to the density jump ratio of 10 (left panel), the other
   corresponds to the jump ratio 100 (right panel).
   The vertical dashed line marks the time of encountering the density jump.
   The green solid lines represent $\Gamma_2$ before encountering the density jump and $\Gamma_4$ during the encountering,
   both denoted by $\Gamma_4$. The blue solid lines and red solid lines represent $\Gamma_2$ (or $\Gamma_3$, $\Gamma_2 = \Gamma_3$),
   and the thermal Lorentz factor of the baryons in Region 3 ($\Gamma_{43}$) after the encountering, respectively. The horizontal dashed lines
   mark the values of corresponding Lorentz factors (which remain constant during the reverse shock crossing time) given by \cite{Dai02}.
   Note that $\Gamma_4$ (see the green solid lines) during the encounter is almost constant (slowly increasing due to adiabatic expansion),
   which is shown as a ``plateau'' that is significantly shorter than that of $\Gamma_2$.
   The shortness of the ``plateau'' of $\Gamma_4$ is due to the different transformation formula between the burst frame time
   and the apparent time in the observer frame.}
   \label{Fig:plot1}
   \end{center}
\end{figure}

\begin{figure}
   \begin{center}
   \includegraphics[scale=0.5]{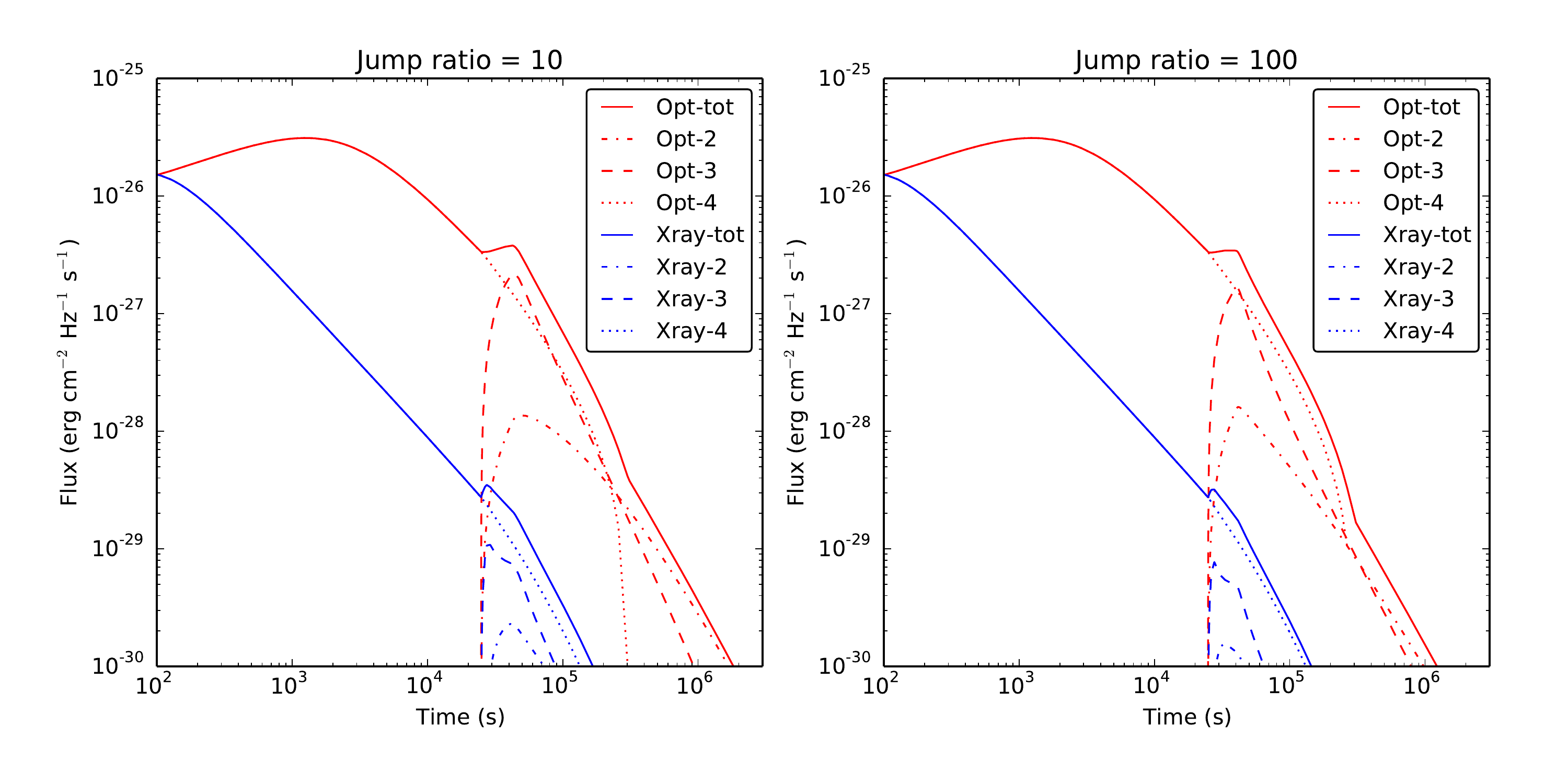}
   \caption{Corresponding afterglow lightcurves for the two cases in Figure 1.
   The red, blue lines are lightcurves in the optical band ($4 \times 10^{14}$ Hz) and the X-ray band (0.3 keV), respectively.
   The dotted lines (Opt-4 or Xray-4) represent the flux density of Region 2 before the density jump and the flux density of Region 4 after the jump.
   The dashed lines (Opt-3 or Xray-3) are the contribution from Region 3 after the density jump.
   The dash-dotted lines (Opt-2 or Xray-2) mark the flux density of Region 2 after the jump.
   Total flux densities of all the components are presented by the solid lines (Opt-tot or Xray-tot).
   For a similar plot, see \cite{Geng14}.}
   \label{Fig:plot2}
   \end{center}
\end{figure}

\section{Delayed Energy Injection Model}
We now focus on the afterglows with steep optical re-brightenings.
We show the steep optical re-brightening can be naturally generated
in the delayed energy injection scenario.
In this model, the central engine after burst is assumed to be a BH.
Considering the vicinity along the spin axis of the BH at late times
should be clean, we can assume the energy flow is in type of Poynting-flux.
The delayed Poynting-flux would be absorbed by the hot plasma behind the FS
and modify the dynamics of the FS.
If the luminosity of the Poynting-flux is $L$,
then the dynamic of the FS can be described by \cite{Geng13}
\begin{equation}
\frac{d \Gamma}{d m} = - \frac{(\Gamma^{2} - 1)-\frac{1 - \beta}{\beta c^{3}} \Omega_{j} L(t_{\rm b} - R/c) \frac{d R}{d m}}{M_{ej} + 2 (1 - \varepsilon) \Gamma m + \varepsilon m},
\end{equation}
where $\Gamma = 1 / \sqrt{1 - \beta^{2}}$ is the bulk Lorentz factor of the FS, 
$\Omega_{j} = (1 - \cos \theta_{j}) / 2$ is the beaming factor of the GRB outflow, 
$m$ is the swept-up mass by the shock, $\varepsilon$ is the radiative efficiency, 
$R$ is the radius of the FS and $t_{\rm b}$ is the time from the event measured in the burster frame.
Such a delayed energy injection would lead to a rapid change
in the evolution of $\Gamma$ according to Equation (3),
consequently the flux would show a steep rise.

The energy injection power (with a luminosity of $L$) during the fall-back accretion may come from some
magnetic processes \cite{Blandford77,Lee00,Yuan12},
however, its exact temporal profile is still uncertain.
Here, we use two possible modes in our calculations.
One is the top-hat mode in which the injected power $L$ is
a constant from a start time $t_{\rm obs}^{\rm s}$ to an end time $t_{\rm obs}^{\rm e}$ \cite{Kumar08}.
The other is the broken-power-law mode, in which the luminosity profile
is similar to the profile of the mass accretion rate during the fall-back
\cite{MacFadyen01,ZhangW08,Dai12,Wu13}, i.e.,
\begin{equation}
L = L_{\rm p} [\frac{1}{2} (\frac{t_{\rm obs} - t_{\rm obs}^{\rm s}}{t_{\rm obs}^{\rm p} - t_{\rm obs}^{\rm s}})^{- \alpha_r s} + \frac{1}{2} (\frac{t_{\rm obs} - t_{\rm obs}^{\rm s}}{t_{\rm obs}^{\rm p} - t_{\rm obs}^{\rm s}})^{- \alpha_d s}]^{-1/s},
\end{equation}
where $L_{\rm p}$ is the peak luminosity at the peak time $t_{\rm obs}^{\rm p}$, $\alpha_{r}$, $\alpha_{d}$ are the rising and decreasing index respectively, $s$ is the sharpness of the peak.
In Figure 3, we show lightcurves calculated in the two modes,
from which we find the steep optical re-brightening would be generated.
The initial conditions of the outflow and the key parameters involved in 
the radiation process are the same as those in Section 2.

This model has been used to interpret the steep re-brightenings in
the lightcurves of GRB 081029 \cite{Nardini11,Holland12,Geng13}
and GRB 100621A \cite{Greiner13,Geng13}.
We notice that $t_{\rm obs}^{\rm s} / (1+z)$ is just equal to $t_{\rm fb}$,
which gives
\begin{equation}
\frac{t_{\rm obs}^{\rm s}}{1+z} \simeq (\frac{\pi^2 r_{\rm fb}^3}{8 G M_{\rm BH}})^{1/2},
\end{equation}
where $r_{\rm fb}$ is the fall-back radius,
$G$ is the gravitational constant, and
$M_{\rm BH}$ is the mass of central BH.
Moreover, the injected energy should come from the
potential energy of the fall-back material, i.e.,
\begin{equation}
\frac{\Omega_j}{1+z} \int^{t_{\rm obs}^{\rm e}}_{t_{\rm obs}^{\rm s}} L d t_{\rm obs} \simeq \eta \frac{G M_{\rm BH} M_{\rm fb}}{r_{\rm fb}},
\end{equation}
where $M_{\rm fb}$ is the total fall-back mass,
$\eta$ is the efficiency of the energy conversion.
In the top-hat mode, Equations (5) and (6) will give
\begin{equation}
M_{\rm fb} = \frac{2}{1+z} \eta^{-1} \left(\frac{t_{\rm obs}^{\rm s}}{1+z}\right)^{2/3} (\pi G M_{\rm BH})^{-2/3}
\Omega_j L (t_{\rm obs}^{\rm e} - t_{\rm obs}^{\rm s}).
\end{equation}
Therefore, in principle, $M_{\rm fb}$ can be inferred from
the fitting to the optical re-brightenings in the scenario.

\begin{figure}
   \begin{center}
   \includegraphics[scale=0.5]{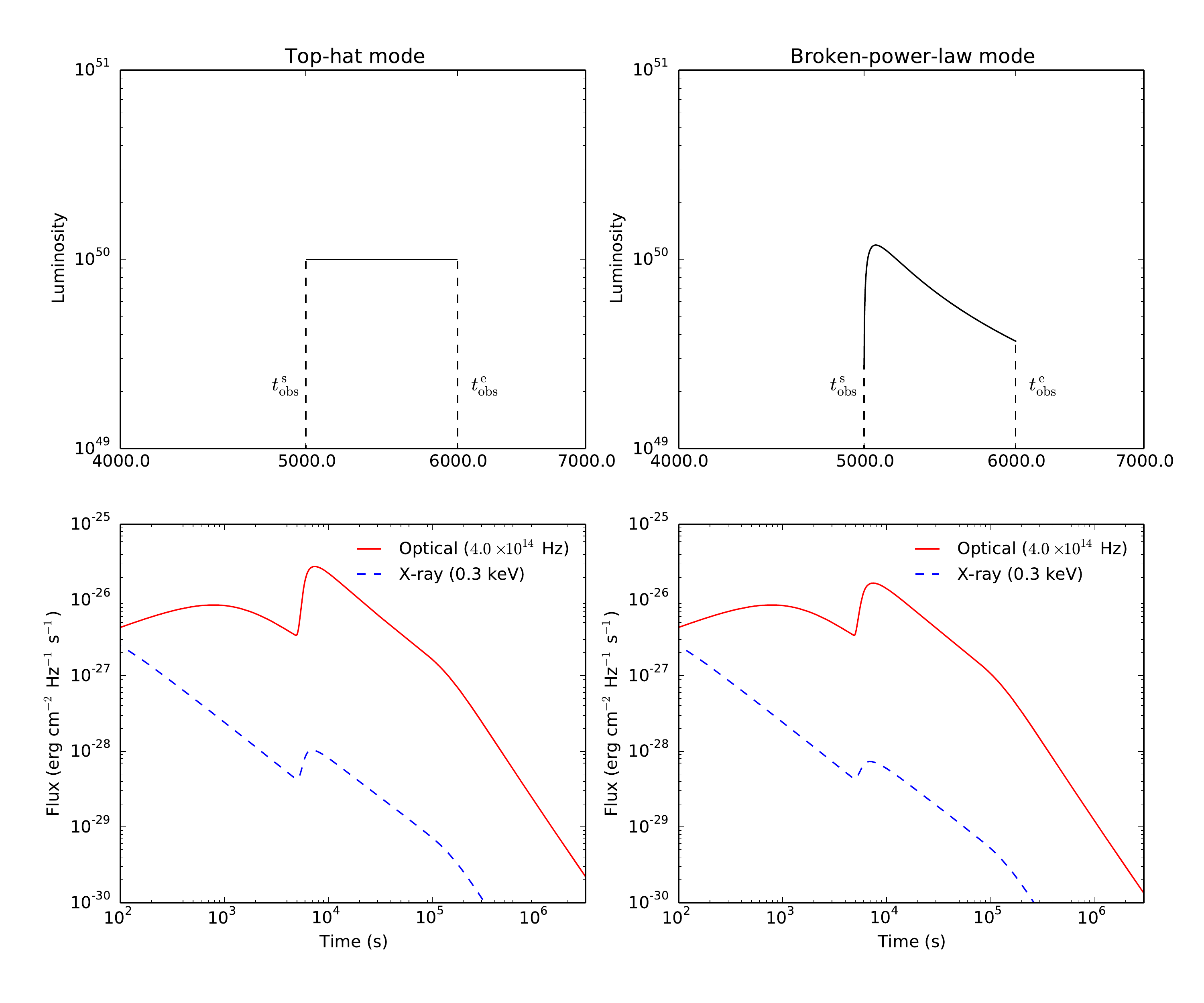}
   \caption{Top left panel: schematic illustration of the top-hat injection mode, in which
   $L_0 = 10^{50}$~erg~s$^{-1}$ (isotropic), $t_{\rm obs}^{\rm s} = 5000$~s, $t_{\rm obs}^{\rm e} = 6000$~s.
   Top right panel: schematic illustration of the broken-power-law (see Equation (4)) injection mode, in which
   $L_{\rm p} = 10^{50}$~erg~s$^{-1}$ (isotropic), $t_{\rm obs}^{\rm s} = 5000$~s, $t_{\rm obs}^{\rm p} = 5200$~s,
   $t_{\rm obs}^{\rm e} = 6000$~s, $\alpha_r = 0.5$, $\alpha_d = -1.5$ and $s = 0.5$.
   The bottom panels show the corresponding afterglows of the two modes correspondingly.
   In the calculations, a redshift of $z = 1$ is assumed.
   Similar plots can be found at \cite{Geng13}.}
   \label{Fig:plot3}
   \end{center}
\end{figure}

\section{$e^+e^-$ Wind Model}
In the prior section, we interpret the steep optical re-brightenings by using
the delayed energy injection model.
Now, we discuss another group of afterglows, of which the
optical re-brightenings are shallower.

After a GRB, the remaining object of the progenitor may be a magnetar,
which will lose its rotational energy by ejecting a continuous Poynting-flux.
In the $e^+e^-$ wind model,
the Poynting-flux may convert into an $e^+e^-$ wind, as hinted from phenomena
associated with pulsar wind nebulae \cite{Rees74,Lyubarsky01,Metzger14}.
As the $e^+e^-$ wind catches up with the FS, a long-lasting
RS will form and propagate back into the $e^+e^-$ wind.
The RS-shocked $e^+e^-$ will act as another emitting source
 besides the electrons shocked by the FS.
Consequently, the afterglow lightcurves are the combination of two components.
This model has been proposed to interpret the X-ray plateau previously.
However, we find this model may account for the common origin for shallow
optical re-brightenings around $10^4$ s.

For a newly born magnetar, its Poynting-flux luminosity $L_{\rm w}(t_{\rm obs})$ is \cite{Shapiro83}
\begin{equation}
L_{\rm w} \simeq 4.0 \times 10^{47} B_{\rm NS,14}^2 R_{\rm NS,6}^6 P_{{\rm NS},-3}^{-4} \left(1+\frac{t_{\rm obs}}{T_{\rm sd}}\right)^{-2} \rm{erg}~\rm{s}^{-1},
\end{equation}
and its spin-down timescale is
\begin{equation}
T_{\rm sd} \simeq 5.0 \times 10^4 (1+z) B_{\rm NS,14}^{-2} I_{45} R_{\rm NS,6}^{-6} P_{\rm NS,-3}^2~\rm{s},
\end{equation}
where $B_{\rm NS}$, $R_{\rm NS}$, $P_{\rm NS}$, $I$,
are the surface magnetic field strength, radius, spin period,
and moment of inertia of the magnetar respectively.
The convention $Q_x = Q/10^x$ in cgs units is adopted hereafter.
We assume the Poynting-flux be converted into $e^+e^-$ pairs,
then the particle density in the comoving frame of the unshocked wind
is $n_4^{\prime} = L_{\rm w} / (4 \pi R^2 \Gamma_4^2 m_e c^3)$,
where $\Gamma_4$ is the bulk Lorentz factor of
the unshocked wind (referred to as Region 4 below).
Unlike the rapid evolution of $\Gamma_2$ in the
delayed energy injection scenario, the evolution of $\Gamma_2$
is shallow due to the smoothly evolving $L_{\rm w}$ here.

The dynamics of the FS-RS system here can be solved by applying the method mentioned in Section 2.
Meanwhile, another method, called the mechanical method \cite{Beloborodov06,Uhm11},
was also proposed to solve the dynamics of such FS-RS system.
Here, we first compare these two methods.
Let's consider an outflow with an isotropic kinetic energy of
$E_{K,\rm{iso}} = 8.0 \times 10^{52}$ erg and
an initial Lorentz factor of $\Gamma_{2,0} = 150$,
and we set $n_0= 0.1$~cm$^{-3}$, $\Gamma_4 = 10^4$,
and $B_{\rm NS} = 2 \times 10^{14}$ G.
The evolution of $\Gamma_2$ can thus be obtained by using the two methods respectively.
In Figure 4, we see the results from our method and
the mechanical method are consistent with each other.
Below, we adopt the mechanical method to solve the
dynamics of the FS-RS system.

After taking the following parameters:
$\epsilon_{e,2} = 0.05$, $\epsilon_{B,2} = 0.01$, $\epsilon_{B,3} = 0.2$,
$\epsilon_{e,3} = 1-\epsilon_{B,3} = 0.8$, $p_2 = 2.1$, $p_3 = 2.4$, 
the corresponding lightcurves can be calculated
(see Figure 5).
In this case, it is clearly shown that the flux from the RS begins
to exceed that from the FS at $\sim 5 \times 10^4$~s,
leading to the emergence of the optical re-brightening and
the X-ray plateau simultaneously.
Due to the effect of the equal arrival time surface \cite{Waxman97,Granot99,Huang07},
the peak time of the optical re-brightening would be
delayed, i.e., larger than $T_{\rm sd}$.
However, this kind of delay in X-rays can be ignored.
As a result, optical re-brightenings are relatively easier to emerge
than ``X-ray re-brightenings''. In other words,
it is often the case that only a X-ray plateau or no equivalent feature in X-rays
accompanies the optical re-brightening.
This property in our model is interestingly consistent with many observations.
In Figure 6, optical afterglows are calculated using
different values of $B_{\rm NS}$.
From these results, it is found that the flux from the
long-lasting RS would account for the shallow optical re-brightening.
The $e^+e^-$ wind model has been applied \cite{Geng16} to
explain the afterglows of GRB 080413B \cite{Filgas11}, GRB 090426 \cite{Nicuesa11},
GRB 091029 \cite{Filgas12}, and GRB 100814A \cite{Pasquale15}.

Since the flux from the RS is sensitive to $L_{\rm w}$ and $T_{\rm sd}$,
the $e^+e^-$ wind scenario provides a useful way to probe the characteristics of newly-born magnetars.
Considering the fact that $L_{\rm w}$ and $T_{\rm sd}$ are 
uniquely determined by the parameter $B_{\rm NS}$, we suggest that 
$B_{\rm NS}$ may be constrained from the fitting to
the observed re-brightenings.

\begin{figure}
   \begin{center}
   \includegraphics[scale=0.5]{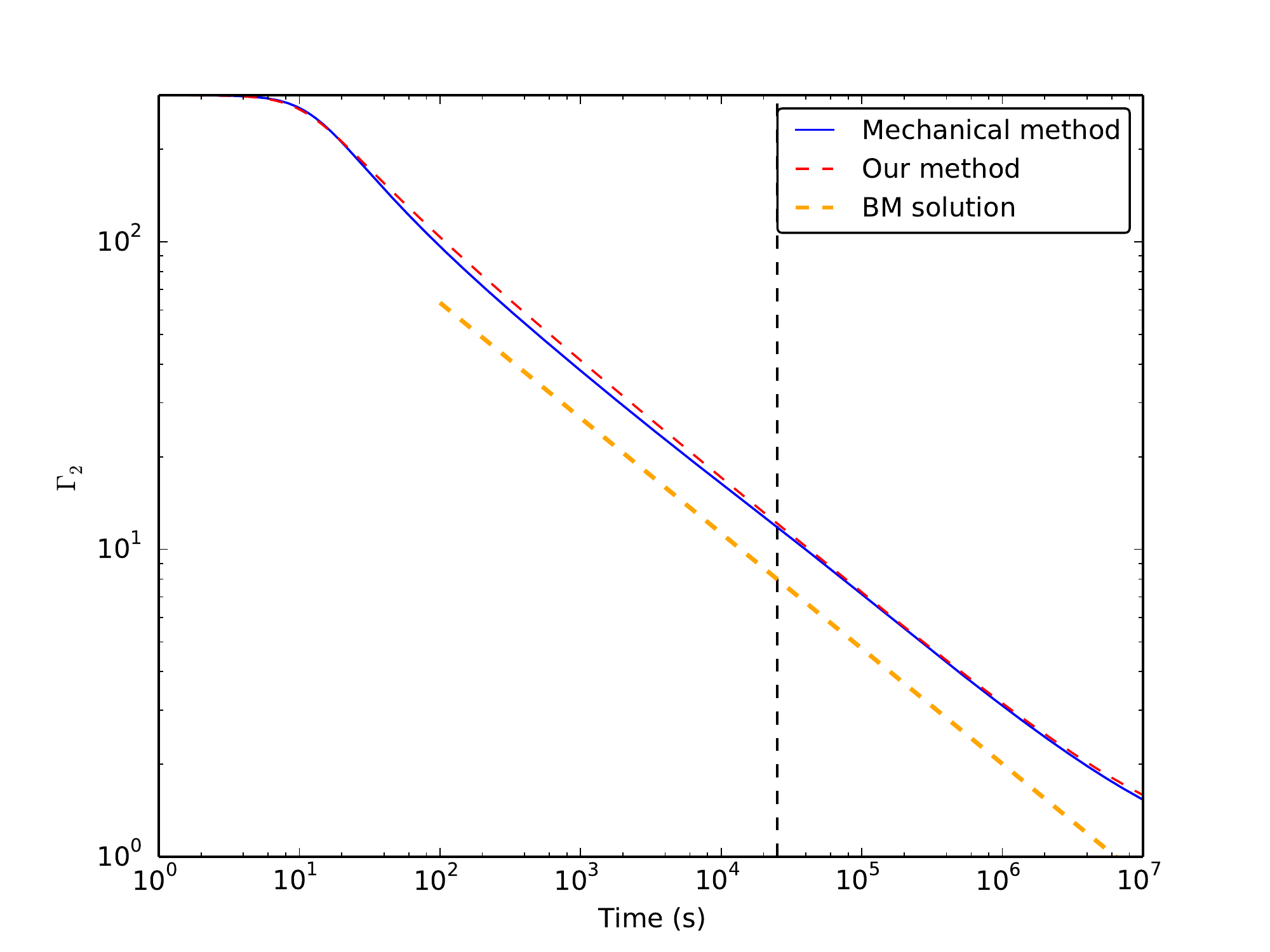}
   \caption{Comparison of the two methods used in solving the temporal evolution of $\Gamma_2$
   in the $e^+e^-$ wind model \cite{Geng16}.
   The initial parameter values are: $E_{K,\rm{iso}} = 8.0 \times 10^{52}$ erg, $\Gamma_{2,0} = 150$,
   $n_0 =0.1$~cm$^{-3}$, $\Gamma_4 = 10^4$, and $B_{\rm NS} = 2 \times 10^{14}$ G.
   The thick dashed orange line represents the BM solution (schematic), i.e.,
   $\Gamma \propto t_{\rm obs}^{−3/8}$ and the vertical dashed line denotes the position of $T_{\rm sd}$.}
   \label{Fig:plot4}
   \end{center}
\end{figure}

\begin{figure}
   \begin{center}
   \includegraphics[scale=0.5]{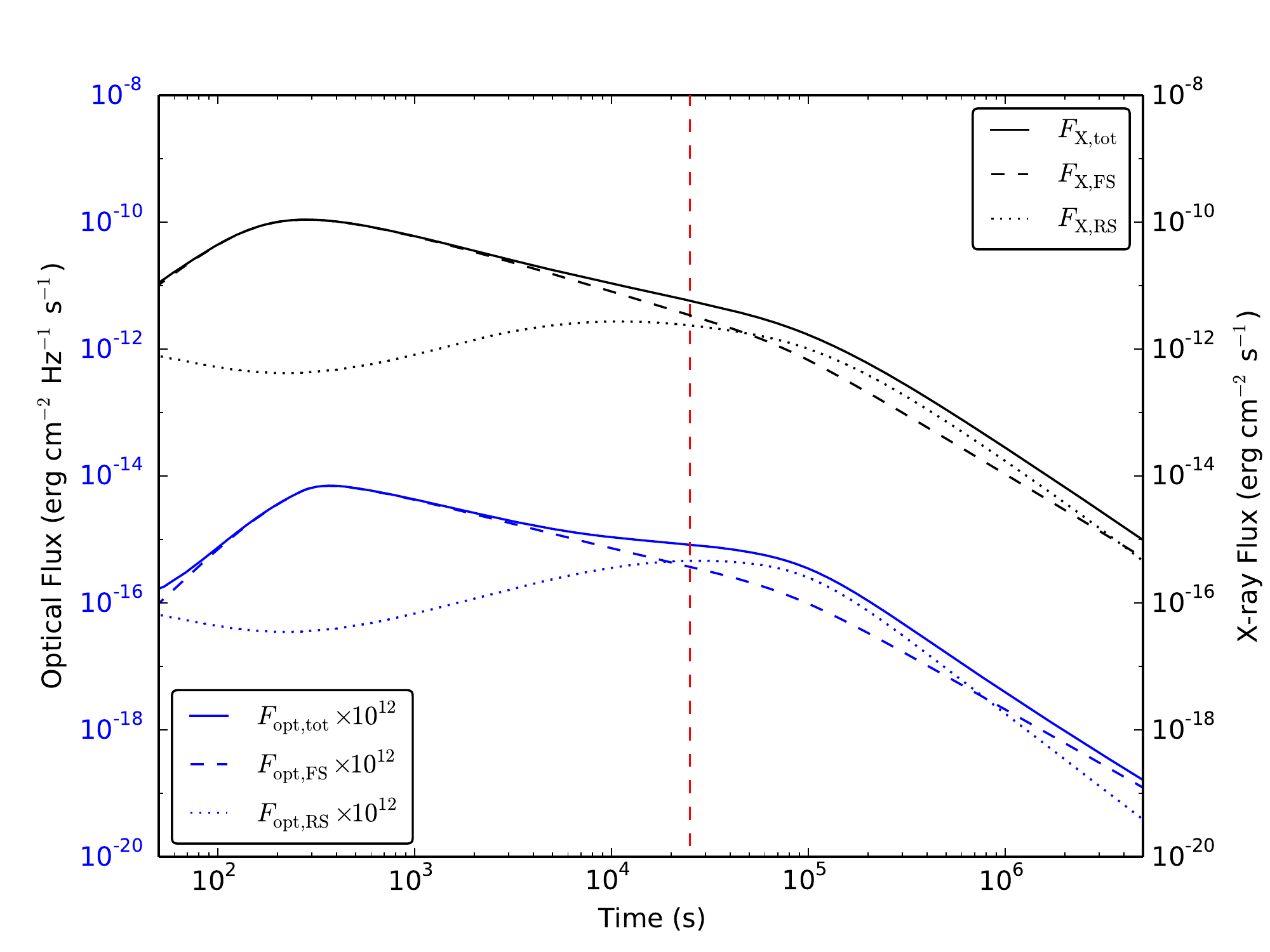}
   \caption{Corresponding afterglow lightcurves of Figure 4.
   The dashed lines represent the flux from Region 3, while the dotted lines
   are emissions from Region 2. The total flux are shown as the solid lines.
   In the calculations, the optical band is taken as $4.0 \times 10^{14}$ Hz and
   the X-ray band is taken as 0.3--10 keV.
   The red dashed vertical line marks the position of $T_{\rm sd}$.}
   \label{Fig:plot5}
   \end{center}
\end{figure}

\begin{figure}
   \begin{center}
   \includegraphics[scale=0.5]{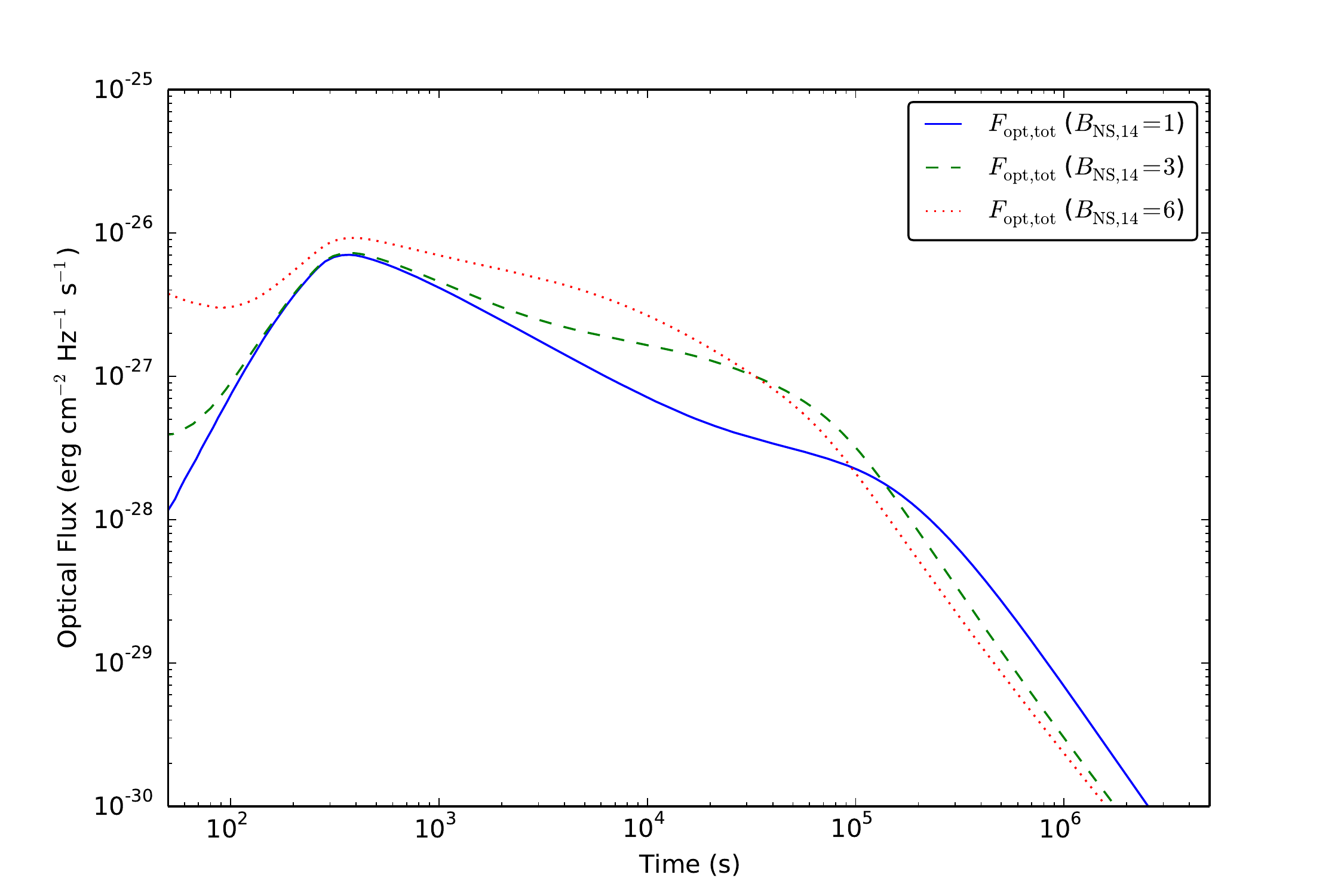}
   \caption{Optical lightcurves in the electron-positron wind model with different $B_{\rm NS}$.
   All the parameters are the same as those in Figure 5,
   except for $B_{{\rm NS}, 14} = 1$ (solid line), $B_{{\rm NS}, 14} = 3$ (dashed line),
   and $B_{{\rm NS}, 14} = 6$ (dotted line).}
   \label{Fig:plot6}
   \end{center}
\end{figure}

\section{Discussion}
In this review, we show that the density jump scenario could not
account for many of the observed optical re-brightenings in GRB afterglows.
Furthermore, we classify the observed afterglows with optical re-brightenings
into two groups and investigate their intrinsic origin.
The afterglows with steep optical re-brightenings are
interpreted by the delayed energy injection model,
which is associated with a central accreting BH.
Meanwhile, the afterglows with shallow optical re-brightenings
are explained by the $e^+e^-$ wind model,
of which the central engine is a magnetar.

Although the above two groups of afterglows have similar origins for re-brightenings,
some other factors will make the situation different.
The observational data of some afterglows clearly show that
their spectral indices are evolving \cite{Filgas12}.
It is hard to satisfactorily model them only by using the simple models discussed here  
and/or other customary models. For the simple $e^+e^-$ wind model or the two-component jet model,
the spectral evolution would occur only when the characteristic frequencies
are crossing the observational band, or when later component begins
to dominate, but it can not work well for some special GRBs.
In such cases, including some microphysical processes
may be necessary to match the observations.
For example, in the varying microphysical parameters scenarios \cite{Kong10b},
the varying electron distribution index $p$ would help to
explain some unexpected spectral evolutions \cite{Filgas12}.
Varying microphysical parameters may be related to
the acceleration performance of relativistic shocks \cite{Sironi13}.
Since the acceleration performance of shocks may depend on
the magnetization (or other factors) of the plasma and the magnetization is 
highly variable \cite{Zhang05}, some special afterglows are foreseeable.

It has been suggested that the two-component jets
could also account for some shallow re-brightenings.
The collapsar model of long-duration GRBs offers
a natural mechanism to generate two-component jets,
i.e., a high speed jet emerging from a star is accompanied by 
a relatively slow cocoon \cite{Ramirez02,Lazzati15}.
The re-brightening lightcurve itself will not help to definitely discriminate
the $e^+e^-$ wind model from the two-component jet model,
since the role of the wide jet is somehow similar to the role of the RS. 
However, there is another way that can help us in
the future. A two-component jet should be associated
with a collapsar. If the re-brightening is observed
to be associated with the double NS merger (by detections of the
gravitational waves \cite{LIGO16}) but not a collapsar,
then the $e^+e^-$ wind model would be preferred.

The intrinsic origins of optical re-brightenings would
help to probe the characteristics of central engines.
In the delayed energy injection model,
$t_{\rm fb}$ can be derived from the start time of the
re-brightening, and $r_{\rm fb}$ can thus be obtained.
If the observational data is good enough,
$L$ can be constrained in the fitting process,
and we can estimate the total mass of the fall-back material.
Furthermore, the fall-back accretion theory requires that
the steep optical re-brightening should be accompanied by
a low energy supernova, in which the fall-back material can
survive during the explosion. This model thus can be
tested by future observations of GRB-supernova association.
In the $e^+e^-$ wind model, $T_{\rm sd}$ can be roughly inferred 
from the peak time of the optical re-brightening.
In general, an earlier re-brightening means
that $B_{\rm NS}$ is larger or $P_{\rm NS}$ is smaller.
So $B_{\rm NS}$ and $P_{\rm NS}$ can be constrained from observations.
Thus the $e^+e^-$ model provides a potential way to probe
the characteristics of the central magnetar.

\section*{Acknowledgement}
We thank Liang Li for helpful discussion.
This work was supported by the National Basic Research Program
of China with Grant No. 2014CB845800, and by the National Natural Science
Foundation of China with Grant No. 11473012.

\end{document}